# Active Material, Optical Mode and Cavity Impact on electro-optic Modulation Performance

Rubab Amin[1], Can Suer[1], Zhizhen Ma[1], Jacob B. Khurgin[2], Ritesh Agarwal[3], Volker J. Sorger[1,*]

[1]Department of Electrical and Computer Engineering, George Washington University
800 22nd St., Science & Engineering Hall, Washington, DC 20052, USA
[2]Department of Electrical and Computer Engineering, Johns Hopkins University, Baltimore, Maryland 21218
[3]Department of Materials Science and Engineering, University of Pennsylvania, Philadelphia, PA 19104, USA
*Corresponding Author: E-mail: sorger@gwu.edu

**Abstract**

**In this paper, three different materials – $Si$, ITO and graphene; and three different types of mode structures – bulk, slot and hybrid; based on their electro–optical and electro–absorptive aspects in performance are analyzed. The study focuses on three major characteristics of electro– optic tuning; i.e. material, modal and cavity dependency. The materials are characterized with established models and the allowed ranges for their key parameter spectra (at $\lambda = 1550\ nm$) are analyzed with desired tuning in mind; categorizing into n and $\kappa$ – dominant regions for plausible electro–optic and electro–absorptive applications, respectively. A semi–analytic approach, with the aid of FEM simulations for the eigenmode calculations, was used for this work. Electro– optic tuning i.e. resonance shift properties inside Fabry–Pérot cavities are investigated with modal and scaling concerns in mind. Tuning changes the effective complex refractive index of the mode dictated by the Kramers–Kronig relations which subsequently suggest a tradeoff between the resonance shift, $\Delta\lambda$ and increasing losses, $\Delta\alpha$. The electrical tuning properties of the different modes in the cavity are analyzed, and subsequently a figure of merit, $\Delta\lambda/\Delta\alpha$ was chosen with respect to carrier**

**concentration and cavity scaling to find prospective suitable regions for desired tuning effects.**

## 1. Introduction

The need for modulation is undoubtedly essential for any switching or computations. As the transition from pure electrical to opto-electronic and photonic devices and architecture are imminent in the near future to enable nanoscale operations and better on-chip packaging density, the need for strong optical modulation is evident. Modulation arises from the light–matter interaction (LMI) within the active material and strong modulation requires active material that exhibits efficient interaction with the light.[1,2] In this context, mainly two types of modulators are relevant – Electro-Optic modulators (EOM) and Electro-Absorptive modulators (EAM). In both types of modulation, the fundamental complex index of refraction is altered in some way for the active material. EOMs operate by changing the real part of the index which relates to the phase of the light, whereas EAMs operate by changing the imaginary part of the index which relates to the intensity absorption of the light. In this work, we primarily focus on the EOMs, but relevant EAM results are also commented on. The challenge for EO modulation is fundamental. Kramers–Kronig relations dictate that only changing the real part of the complex index is impossible. As the real part changes, the imaginary part does the same also. Usually the real part decreases with modulation and simultaneously the imaginary part increases, this affects the performance of the device as a rise in the imaginary part of the index imposes loss in the material.

The active material changes its optical characteristics, i.e. the real and imaginary parts of the index, with modulation. Conventional electro-optic (EO) materials eg. Silicon, which has shown Kerr, and carrier injection effect, has rather low modulation strength, due to the small refractive index change.[3,4,5] The use of nanometer thin materials expanded the horizons of material sciences and this led researchers to use materials like transparent conducting oxides (TCOs), which can operate in higher ranges in the electromagnetic spectrum. The most common among TCOs, in application, is Indium Tin Oxide (ITO). ITO is an emerging EO material which has unity order change in refractive index under electrical gating.[6,7] The drastic change in graphene's refractive index and extinction coefficient make it a suitable material for both EO and EA (electro-absorptive) operations. Graphene has shown electro-optic response via Pauli-blocking in near IR frequencies and modulating functionality.[8,9] The strong change in the real refractive index is due to the strong effect through Pauli blocking, thus making graphene a naturally suitable material for EOMs. But the atomic thin thickness of single layer graphene produce a challenge for modulation because of the miniscule optical confinement factor and the in-plane electric field selectivity of graphene for LMI. However, exploiting the drastic change of the index, efficient modulation can be obtained using suitable modal structures. Carriers begin to accumulate on the graphene sheet when a voltage is applied to the graphene/oxide capacitor, and thus the chemical potential, $\mu_c$ is tuned with the gate voltage. As $|\mu_c|$ reaches half of the equivalent photon energy ($1/2\ h\nu$), the interband transition is blocked hence the absorption greatly decreases.[10]

Different mode structures confine light in different ways and dimensionalities. The bulk photonic modes are often leaky as LMI is little and require more real estate on chip. To enhance LMI and enable nanoscale operation plasmonic modes supporting surface plasmon polaritons (SPP) can be employed. Plsmonic modes can offer more index change and hence more modulation strength as compared to bulk modes, but plasmonic modes are inherently quite lossy. Using plasmonic modes in a system with feedback gain thus automatically encounters tradeoff, which naturally suggests an optimum length of the device in question where the gain and losses are in equilibrium. Bulk photonic modes can offer more active material to contribute to index modulation, but even with diffraction limited dimensions, need larger footprint for essential performance. SPPs however can offer footprint friendly dimensions, but incur substantial loss to factor into the performance. The device performance of EOMs greatly depend on a variety of interrelated design concepts; the underlying waveguide determining coupling and propagation losses, the confinement factor of the active material with the optical mode, the strength of the optical index change being altered,[6,11] and subsequent impacts on energy efficiency, modulation speed, footprint, and optical power penalty.[8,12–15] Previous work focused on addressing these in an add-hoc manner.[6,16,17]

Enhancing LMI is the main target to get better modulation strength. This can be achieved by the use of optical cavities as the feedback provided by the cavity essentially can be interpreted as folding back the optical signal additional times inside the device to interact further with the active material to contribute to the index modulation. This effect will necessarily be proportional to the Q–factor of the cavity. Using optical cavities allow enhancing the interaction strength of light with the active material, but such approaches can limit the energy-per-bit efficiency, which scales proportional with device volume.[18] While a longer cavity is expected to provide higher Q, but inherently lossy material with lossy mode combinations suggest a tradeoff there as well, which can lead to optimal scaling considerations. In this work, we consider Fabry–Perot cavities based on our choice of material/mode combinations to outline EO modulation – loss relations and optimal scaling.

In modulator work, the typical nomenclature regarding the states of operation for the device relates to the light transmission characteristics through the device, rather than states based on voltage bias. Modulator on state implies the state where transmission is highest, which corresponds to the low loss state. Similarly, modulator off state signifies the lossy state corresponding to a drop in transmission. In this work, we use this terminology to characterize our operational states.

## 2. Kramers – Kronig Relations in Tuning

Kramers–Kronig dispersion relations are fundamental with respect to the light matter interaction (LMI) phenomena in transparent matter, molecules, gases, and liquids. They provide restrictions for ensuring self-consistency of experimental or model-generated data. Also Kramers–Kronig relations allow for the possibility for optical data inversion, i.e. information on dispersive phenomena can be obtained by converting measurements of absorptive phenomena over the whole spectrum and vice versa. Especially, these general properties permit the framing of distinguishing phenomena that are relevant at given frequencies showing that their dispersive and absorptive contributions are connected to all the other contributions in the rest of the spectrum.[19] Such general properties relate to the principle of causality in light-matter interaction. The principle derives from time order of cause and response of a system; that is, it states that the effect cannot precede the cause.[20] The passage of electromagnetic waves through dielectric media is characterized by a complex permittivity, leading to a complex refractive index,

$$\tilde{n}(\omega) = n(\omega) - i\kappa(\omega) = \sqrt{\tilde{\epsilon}(\omega)} = \sqrt{\epsilon'(\omega) - i\epsilon''(\omega)} \qquad (1)$$

where $\omega$ is the angular frequency, $n(\omega)$ is the real part of the refractive index, $\kappa(\omega)$ is the imaginary part of the refractive index and $\epsilon'$ and $\epsilon''$ presenting the real and imaginary parts of the permittivity, respectively. The basic principle of causality of the response of any material to an external electromagnetic field ensure that the permittivity being a holomorphic function in the upper half of the complex frequency plane, obeys Kramers–Kronig relations that connect the real and the imaginary parts of the refractive index [19,21–23]

$$n(\omega') - 1 = \frac{2}{\pi} P \int_0^\infty \frac{\omega\kappa(\omega)}{\omega^2 - \omega'^2} d\omega \qquad (2)$$

$$\kappa(\omega') = -\frac{2\omega'}{\pi} P \int_0^\infty \frac{n(\omega)-1}{\omega^2 - \omega'^2} d\omega \qquad (3)$$

where $P$ signifies that the principal value of the integral is taken at the point $\boldsymbol{\omega = \omega'}$. The Kramers–Kronig relations can model the propagation of light in a homogeneous and lossy dielectric medium relating the real and imaginary parts of the index to each other. With the aid of traditional material models, it is established that with tuning the material changes its optical properties. Tuning or modulation causes the real index, $n$ to drop. Kramers–Kronig relations dictate that in order for the real part to drop, the imaginary part i.e. the extinction coefficient, $\boldsymbol{\kappa}$ must rise. Consequently, electro-optic tunability of materials or resonance shift in cavities and associated losses thereof are competing factors which are related by the Kramers–Kronig relations, as one cannot have an ideal active material that has only a real or imaginary index change.

## 3. Material Models & Electrical Tuning

### 3.1. The Drude Model for Silicon and ITO

The active material changes its optical properties with modulation. There are established models that characterize the change in these properties. For Silicon and ITO, we characterize the corresponding complex refractive index change from electrical tuning with the Drude model. The permittivity $\tilde{\epsilon} = \epsilon' - i\epsilon''$, as defined by Drude model, is given by

$$\tilde{n}^2 = \tilde{\epsilon} = \epsilon_\infty - \frac{\omega_p^2}{\omega(\omega + i\gamma)} \qquad (4)$$

where $\epsilon_\infty$ is the high-frequency dielectric constant, $\omega$ is the angular frequency of the illuminating light, $\gamma = 1/\tau$ is the carrier scattering rate i.e. collision frequency, and $\omega_p$ is the unscreened plasma angular frequency defined by $\omega_p^2 = N_c q^2 / \epsilon_0 m^*$.[24,25] Here $N_c$ is the carrier concentration, q the electronic charge, $\epsilon_0$ the permittivity of vacuum, and $m^*$ the effective carrier mass The electron mobility $\mu$ and $\tau$ are related by $\mu = |q|\tau/m^*$.

Silicon is a conventional photonic material for ease of fabrication and pricing concerns. In this work to characterize the complex refractive index change with EO modulation, we use the dielectric constant of undoped silicon $\varepsilon_\infty = \varepsilon_{Si} = (3.415)^2$ for the Silicon Drude model computations.[26] The effective mass, $m^*$ taken as $0.26m_0$, where $m_0$ is the free electron mass,[27] $\mu$ is taken as 1100 cm$^2$V$^{-1}$s$^{-1}$.[24] As the drive voltage is increased, more free carriers are created which leads to dispersion in the first place. This effectively increases the carrier concentration in the material. For EO modulation, the dispersion relations were calculated by varying carrier concentration, $N_c$ in the permitted range it follows i.e. $10^{16}$–$10^{20}$ cm$^{-3}$.

In the near-infrared, ITO responds to light as a metal would, indicating that its free electrons dictate its optical response. In fact, ITO and related TCOs have recently been explored as plasmonic materials in the optical frequency range.[28–33] It has been shown that the Drude model predicts the permittivity of ITO at wavelengths beyond 1 μm well.[25] Several previous studies have calculated the permittivity of ITO using the experimentally measured reflectance and transmittance, and we choose a fitting result of Michelotti et. al whereas $\epsilon_\infty$ and γ depend on the deposition conditions, in our analysis we have taken $\epsilon_\infty = 3.9$ and $\gamma = 1.8 \times 10^{14}$ s$^{-1}$.[25,29] The plasma frequency, $\omega_p$ is mainly determined by the ITO doping concentration, $N_c$, which is in the range of $10^{19}$–$10^{21}$ cm$^{-3}$ depending on the deposition conditions, defect states, and film thicknesses.[30,32] In this study, we take the doping $N_c$ $=10^{19}$ cm$^{-3}$ for the EO modulator on state.

### 3.2. The Kubo Model for Graphene

Graphene shows anisotropic material properties given its dimensions: in its honeycomb like lattice plane, the in-plane permittivity ($\epsilon_{||}$) can be tuned by varying its chemical potential $\mu_c$, whereas the out-of-plane permittivity is reported to remain constant around 2.5.[34] In this work, graphene is modeled by a surface conductivity model $\sigma(\omega, \mu_c, \gamma, T)$, where $\omega$ is the angular frequency, $\mu_c$ is the chemical potential, $\gamma$ is the scattering rate and T is the temperature. The conductivity of graphene is given by the Kubo formula:[35]

$$\sigma(\omega, \mu_c, \gamma, T) = \frac{iq^2(\omega - i2\gamma)}{\pi\hbar}\left[\frac{1}{(\omega - i2\gamma)^2}\int_0^\infty \varepsilon\left(\frac{\partial f_d(\varepsilon)}{\partial\varepsilon} - \frac{\partial f_d(-\varepsilon)}{\partial\varepsilon}\right)d\varepsilon - \int_0^\infty \frac{f_d(-\varepsilon) - f_d(\varepsilon)}{(\omega - i2\gamma)^2 - 4(\varepsilon/\hbar)^2}d\varepsilon\right] (8)$$

where $f_d(\varepsilon) = 1/\left[e^{(\varepsilon - \mu_c)/k_B T} + 1\right]$ is the Fermi-Dirac distribution function, $\hbar$ is the reduced Planck's constant $\hbar = h/2\pi$, and q is the electron charge. With applied gate voltage, the chemical potential $\mu_c$ can be tuned, thus the optical property of graphene is a function of gate voltage, $V_g$ and the capacitor model, which is quite similar to the ITO case. We can calculate the in-plane conductivity $\sigma_{||}$ from the Kubo formula. Then the in-plane permittivity is calculated by $\varepsilon_{||} = 1 - \frac{\sigma_{||}}{i\omega\varepsilon_0\Delta}$ where $\Delta = 0.35$ nm, is the thickness of graphene sheet, and the refractive index and extinction coefficient are calculated from the complex in-plane permittivity.[10] Unlike bulk materials, the unique electro-optic properties of graphene dictate its modulation response in a manner such that the out of plane components of the electric field do not contribute to any change in the material, only the in-plane components correspond to enhance LMI.

## 4. Steps in Tuning

### 4.1. Material Tuning Properties

Towards evaluating the EO modulation potential of the active material itself, here we investigate the material aspects of tuning for our chosen materials, i.e. Silicon, ITO and Graphene. The Silicon Drude model is used to calculate the refractive index and extinction coefficient dispersion relations with respect to various wavelengths for a range of permitted carrier concentrations from $10^{16}$ cm$^{-3}$ to $10^{20}$ cm$^{-3}$. In this work we assume a fixed operating wavelength at 1550 nm, corresponding to the telecom C band, and derivatives are taken for the real and imaginary components of the complex refractive index. Starting with Silicon, from **Figure 1a, 1b** it is apparent that the real part of the index, $n$ changes notably with carrier concentration. However, while the relative change of the imaginary part of the index, $\kappa$, appears significant, the small absolute value of the OFF state (high carrier concentration) does not make Silicon a high performing EAM material. This shows that Silicon may be used for its EO properties, but it is not well suited for EA operations. This becomes evident when plotting $\delta n / \delta \kappa$; for higher carrier concentrations the slope kinks downwards for carrier concentrations above $10^{19}$ cm$^{-3}$ (Figure 1h) which indicates the material becoming more lossy, but not $\kappa$ – dominant or favorable for EA operations yet. Also, the gradual slope in $\delta n / \delta \kappa$ suggests weak EO modulation before this transition region making the entire carrier concentrations range $n$ – dominant, which is analogous to favoring EO operations.

The Drude model characterizes the ITO material accurately within our specific wavelengths of interest (NIR regime). Modulation is an effect of change in the refractive index; and the broadening of the corresponding resonances inside a cavity with respect to wavelength can be related back to the loss in the material absorption itself, which corresponds to the extinction coefficient, $\kappa$. When ITO is packaged as one electrode of an electrical capacitor, applying the voltage can put the system into the three known states of accumulation, depletion, or inversion. For example, in accumulation, free carriers are accumulated in the interface of the ITO and the oxide, thus changing the carrier concentration. From the Drude model, explicit relations for both the index and extinction coefficient as functions of the carrier concentration can be obtained. The optical property of the active material therefore changes dramatically, resulting in strong optical modulation effects.

Derivatives of the real and imaginary parts of the complex index with respect to the carrier concentration are taken, from whom the respective derivative $\delta n / \delta \kappa$ can be calculated. It is interesting to note that the ratios $\delta n / \delta N_c$ and $\delta \kappa / \delta N_c$ intersect each other close to the epsilon near zero (ENZ) region at a carrier concentration of $6.2 \times 10^{20}$ cm$^{-3}$. The relative change in $\delta \kappa / \delta N_c$ is more than the relative change in the $\delta n / \delta N_c$ after a certain point in the carrier concentration prior to ENZ, which suggests the effect of $n$ change with modulation is stronger to the left of this point and the effect of $\kappa$ change is stronger to the right. This can be observed in the graph for the derivatives with respect to the carrier concentration for ITO (Figure 1j) This phenomenon suggests that the material is $n$ – dominant to the left of this point, and $\kappa$ – dominant to the right. The ITO material can be used for the EO case in the $n$ – dominant regime, and for the EA cases in the $\kappa$ – dominant regime. The material can be termed $n$ or $\kappa$ – dominant, based on the carrier concentration regions where the change in one exceeds the change in the other corresponding to tuning.

The graphene dispersion relation can be obtained via the Kubo formula (Equation 8, Figure 1e, 1f). The material $n$ and $\kappa$ for graphene are calculated with varying wavelength and chemical potential. We note that the chemical potential, $\mu_c$ for graphene used in the Kubo formula is equivalent to the carrier concentration in the Drude model, and used as the voltage-dependent index tuning parameter. Due to the unique electro-optic property of graphene, $n$ changes abruptly for $\mu_c$ values between 0.4 eV to 0.5 eV, while $\kappa$ changes everywhere else. Thus, 0.4 eV to 0.5 eV is the $n$ – dominant region and elsewhere in the $\mu_c$ range it is $\kappa$ – dominant (shaded region, Inset Figure 1l). Since the $n$ change is abrupt in the aforementioned range of $\mu_c$, and $\kappa$ does not change noticeably $\delta n / \delta \kappa$ has an abrupt peak in that region, resembling a delta function. Due to the unique property of graphene, it is better to characterize the $n$ and $\kappa$ – dominance in the dispersive relation keeping the wavelength fixed at 1550 nm and changing the chemical potential (Figure 1l). Here the $\kappa$ – dominance for graphene is found to the left and right side of the $n$ – dominant region. For the subsequent modal and cavity discussions below we only consider the $\kappa$ –dominance on the left side of the $n$ – dominant region, because we can choose either $\kappa$ – dominant region for EA purposes and the $n$ – dominant region is of interest in EO modulation.

## 4.2. Modal Tuning Properties

Having established the index character regimes as a function of voltage tuning, we are next interested in the possible modal impact of such modulation. In this work, we study three different mode structures for each of the three active materials introduced above (**Figure 2**). Our aim is to explore modulator-suitable material/mode combinations for both EA and EO modulation mechanisms. We aim to explore increased LMIs towards ultracompact modulators while preserving extinction ratio (ER), i.e. modulation depth. Here we considered plasmonics as a spatial mode compression tool towards increasing the LMI and compare two distinct plasmonic modes with a bulk-case for comparison. The two plasmonic modes analyzed are the slot waveguide in a metal-insulator-metal (MIM) configuration,[36–41] and a hybrid photon plasmon (HPP) design in a metal-insulator-semiconductor (MIS). In order to understand the LMI enhancement effect from modal compression, we compare each active material with a bulk case where the entire waveguide consists of the active material only. The resulting design space is a 3×3 matrix, and we track the modal field distribution (Figure 2) as well as effective index for each waveguide case (**Figure 4**).

All mode structures are chosen on top of a $SiO_2$ substrate thus providing a leveled playing field. The bulk $Si$ waveguide is the classical silicon on insulator (SOI) rib waveguide. Starting our discussion with the active material $Si$ case, the height and width of the waveguide were chosen to coincide with the diffraction limited dimension $\lambda/n$ towards maximizing packing density. For all three $Si$ cases we assume that it is possible for the entire $Si$ portion to change index with applied bias. This is possible as the method of modulation is bulk carrier injection as previously demonstrated.[42–46] It is noticeable that this mode is not strongly confined to the physical cross-sectional area of the waveguide mode due to modal field leakage (Figure 2a'). Similarly, the bulk ITO structure is a waveguide made of ITO on top of a $SiO_2$ substrate (Figure 2d'). Index modulation for bulk ITO can in principle done via thermo-refractive effects but the effect is rather slow. The carrier concentration in ITO increases with annealing temperature.[47,48] A second mechanism for index tuning ITO is via a capacitive carrier modulation-based mechanism.[2,6,10,16,38,49,50] Thus, for biasing purposes a 5 nm gating oxide layer is included. Here the accumulated carriers shift the plasma dispersion via the Drude model. In praxis, a 1/e decay length of about 5 nm was measured before,[38] and high index modulation has been experimentally verified over $1/e^2$ (10 nm) thick films.[50,51] As such the modulation mechanism for ITO considered here is capacitive-gated carrier accumulation rather than injection as for the $Si$ case. However similar to the $Si$ bulk case, the bulk ITO waveguide is also rather leaky and unconfined due to the relatively small index contrast with the surrounding air (Figure 2d'). Only the fine accumulation layer of about 10 nm changes properties with tuning while the remainder of the ITO does not change with applied bias. The bulk graphene structure is made by placing a single layer graphene on a gate oxide on top of a *Si* waveguide, thus forming an electrical capacitor. The *Si* waveguide was chosen to have a height of 200 nm, which supports the 2nd order TM mode resulting in an improved modal overlap with the active graphene sheet.[8] The silicon width is diffraction limited $\lambda/n$ similarly as before. It is important to point out that all the waveguides requiring a gate oxide in this work have been designed with a fixed oxide thickness such that we can compare them in a similar standard. That is, to ensure the same electrostatic potential during the comparison, all gate oxides in this work are 5 nm in thickness. The *Si* slot is inspired by the dielectrically loaded surface plasmon polariton (DLSPP) structure; with a 30 nm *Si* layer on top of the substrate is chosen to enhance the plasmonic modal interaction with the *Si* material.[40] This narrow thickness allows the necessary index contrast between the *Si* and *SiO₂* substrate to squeeze the light in the plasmonic gap. Two metal pads are placed to facilitate the plasmonic mode in the gap of 20 nm. From simulation the gap mode is chosen and it is clear that despite being a *Si* slot, a material with almost no inherent loss, the mode is rather lossy and the field intensity in the slot is about 232 times more compared to the bulk case. This is due to the fact that the light is confined in the gap in the slot between the metal walls more than the *Si*. The ITO slot follows the PlasMOStor design.[39] The gap between the metal pads is chosen as a 5 nm *Al₂O₃* layer for gating and subsequently topped with ITO. The slot gap is taken to be 300 nm, for preserving the photonic mode within the gap when ITO is not in the lossy state (modulator ON state). Due to the fact that carrier accumulation can only be around 10 nm, the width of the slot beyond 20 nm will have little impact on confinement, since only activated ITO contributes to the confinement factor with tuning. Our results show a nice confined structure (Figure 2e'). Also, the considerably larger dimension reduces the fabrication complexity. The graphene slot consists of placing a single layer of graphene on top of the *SiO₂* substrate separated by a gating oxide of 5 nm. Then two metal pads form the slot structure. Here the gap is also 20 nm similar to the *Si* slot, which we have found previously to deliver high modulation performance.[10] We note that broader gap dimensions lead to higher order modes, lower optical confinement, and hence lower ER. This value (20 nm) can be understood from two aspects both relating to the fact that metallic confinement beyond 20 nm is not favorable: a) the skin depth of plasmons at telecomm wavelengths is about 20 nm, and b) the Purcell factor reduces dramatically beyond 10's nm small plasmonic cavities due to high losses and field leakage.[18] Our results indeed confirm a modal confinement to the gap and a high field strength, which is 4643 times higher compared to the bulk case (Figure 2h'). Finally, for confinement of light in the transverse direction and to obtain a dielectrically aided plasmonic mode for better modulation control, we choose the hybrid mode structures.[16,52–54] The hybrid *Si* mode is comprised of a metal layer and an oxide layer on top of a *Si* waveguide that is 200 nm thick. The results show a confined mode with most of the light in the oxide. The hybrid ITO mode is similar to the hybrid *Si* mode, with an added 10 nm layer of ITO between the oxide and *Si* layers.[51] The hybrid ITO structure is essentially an MOS structure with a 10 nm ITO layer inserted between the oxide and *Si* interfaces. With the metal on top and *Si* on bottom, it creates an optical capacitor. The gap (oxide + ITO) provides the means to store electromagnetic energy, leading to sub–wavelength optical guiding.[54] The strong energy confinement in the gap arises from the continuity of the displacement field at the material interfaces, which leads to a strong normal electric-field component in the gap.[55] The dielectric discontinuity at the semiconductor–oxide interface produces a polarization charge that interacts with the plasma oscillations of the metal–oxide interface; that is, the gap region has an effective optical capacitance. The hybrid graphene mode is made of a metal layer on top of a 10 nm oxide layer, and the graphene single layer is sandwiched inside the oxide. These are stacked on top of a *Si* waveguide with 200 nm thickness.[6] Both the hybrid ITO and graphene structures show reasonably high confined modes and field enhancements of 46 and 63 times compared to their corresponding bulk cases, respectively. It is worthy to mention that the slot and HPP modes are comparably lossy without accounting for the material loss to contribute as a byproduct of tuning. As such one intuitively would expect these to be suitable for EAM devices. However, there are also regions where the EOMs  (real part index tuning) via phase shifting outperform bulk cases despite the high losses, as discussed below. Also it is noteworthy that, all the metal used for the modal simulations is Gold (*Au*), which has a reasonably low ohmic loss.

Key to the modulator performance such as extinction ratio (ER) is the amount of index change that can be obtained upon biasing the device. Electro–optic modulation is necessarily phase modulation, i.e. the effective change in the k–vector of the light; $\delta k = \delta \omega_0 \frac{\partial k}{\partial \omega} = \frac{\delta n}{n} \omega_0 \frac{\partial k}{\partial \omega} = \frac{\delta n}{n} k_0 c \ \frac{\partial k}{\partial \omega} = \delta n \ k_0 \frac{n_g}{n}$ , where $n_g$ is the group index in the corresponding structure. The phase change then becomes $\Delta \phi = \frac{2\pi}{\lambda} \Delta n_{eff} L = \Gamma \ \delta n \ k_0 \frac{n_g}{n} L$, where $\Gamma$ is the optical confinement factor. The index change inside the modulator ($\Delta n_{eff}$) is then given by $\Delta n_{eff} = \Gamma \ \delta n \frac{n_g}{n}$. For discrete modulation states this relationship can be expressed as the ratio of the altered active material index relative to its initial condition ($\Delta n_{mat}/n_{mat}$) multiplied by its modal confinement factor ($\Gamma$) and effective group index ($n_g$), i.e.[56]

$$ER \propto \Delta n_{eff} = \Gamma \frac{\Delta n_{mat}}{n_{mat}} n_g \tag{11}$$

Note, that the effective group index $n_g$ is inside the cross-sectional structure corresponding to propagation in the longitudinal direction given by $n_g = n_{eff} - \lambda \frac{\partial n_{eff}}{\partial \lambda}$, $\Gamma$ is the optical confinement factor. Equation (11) is true for the isotropic index materials i.e. Silicon and ITO based structures in this work. Due to the unique electro-optic nature of graphene and anisotropy of the indices, this simple equation is not enough to describe modulation performances in the graphene based structures. The graphene propagating energy index and group index need to be represented by directional tensor terms and solved for each component, which is beyond the scope of this work. Here we follow a similar approach for the graphene based modes to the bulk cases in order to relate modulation effects relating to the modal illumination pattern and effective index change. The material index ratio change from Eqn (11) was previously discussed in Figure 1. Next, we focus on the confinement factor first, then discuss obtainable effective index changes that govern the modulator operation. Our modulator bias analysis is based on selecting discrete bias points for all nine waveguides to obtain their corresponding effective indices (Figure 3&4) [SOM ref].

### 4.2.1. Confinement Factor and Index Tuning:

For waveguide modes containing non-diffraction limited mode (bulk) the modal confinement factor is the spatial field (**E**) ratio of the modal overlap of the active region, i.e. the material whose index is electrically being altered, relative to the size of the entire mode:

$$\Gamma_{Bulk} = \frac{\iint_{Bulk} |\mathbf{E}|^2 dS}{\iint_S |\mathbf{E}|^2 dS} \tag{12}$$

Similarly, 2D materials show their unique electro-optic tunability when the electric field is in the lattice plane (i.e. in-plane), but not perpendicular to it (out-of-plane), due to the low polarizability in this direction.[10] Hence, the confinement factor to characterize the light–2D material interaction is given by

$$\Gamma_{2-D} = \frac{\iint_{2-D} |\mathbf{E_{in}}|^2 dS}{\iint_S |\mathbf{E}|^2 dS} \tag{13}$$

where $\mathbf{E_{in}}$ denotes the in-plane electric field, and $\mathbf{E}$ is the electric field. The latter is a critical requirement for 2D materials, since their polarizability vanishes for out-of-plane fields. Thus, simply placing a 2D material on a plasmonic waveguide surface will not lead to increased overlap factors despite the high field concentration near that metal-dielectric interface. In fact, this is a fundamental challenge of interfacing 2D materials with plasmonics. A possible waveguide design, however, to overcome this bottleneck is a MIM-like slot waveguide with a graphene layer parallel to the slot field (Figure 2h).

Starting with the Silicon as the active material, the bulk mode has an overlap factor approaching unity (80%) for the modulator ON-state at low carrier concentrations (**Figure 3a**). However upon plasma dispersive biasing the overlap factor slightly decreases since the mode becomes more lossy. The two plasmonic modes, hybrid and slot (Figure 2b, 2c), worsen the overlap since the Silicon can only sit near the highest fields of both modes, but not directly in it. The latter could be different, for instance, if the slot would be filled with Silicon, but this is not part of this study. The *Si* slot has the lowest confinement of about 20% since the field is mainly concentrated inside the slot. All three Silicon-based waveguides have relatively flat confinement factors with tuning, since the modal distribution does not experience any significant change.

Both the confinement factor and material index change impact the effective index with applied bias (**Figure 4a**). The change in effective *n* is not dramatic but $\kappa_{eff}$ varies rather rapidly with the carrier concentration for the bulk *Si* case. However, even with the increase in $\kappa_{eff}$, it is noticeable that the mode is relatively lossless compared to the other modes studied as *Si* material inherently exhibits little loss. The change in the effective index and effective extinction coefficient follows the material model from previous sections, since this bulk mode is material homogeneous. This is further evident from the confinement factor of the mode; while we find a high modal confinement factor (i.e. good overlap), its value is not changing significantly with increased carrier concentration. The slot and hybrid structures for *Si*, exhibit a higher amount of loss than the bulk *Si*. The change in the real part of the effective index with modulation, in these cases, are stronger than the bulk; as expected from better confined structures. Although the effective extinction coefficients for the slot and hybrid *Si* are relatively flat (Figure 4b), in fact, they do decrease insignificantly towards the end of the permitted carrier concentration range, near $10^{19}$–$10^{20}$ cm$^{-3}$. However, this does not violate the Kramers–Kronig relations because the more tuning is in effect, the more light is projected out from the active *Si*, which becomes more and more lossy with tuning. This effect can be seen from Figure 3, the slot and hybrid *Si* confinement factors decrease towards the right end. This means with tuning the light is squeezed more into the oxide for the hybrid case, and gap for the slot.

Unlike the Silicon case, the three ITO-based waveguides experience a rather strong index change with bias. This is mainly driven by the drastic change in the confinement factor. Starting with the slot waveguide, the field is somewhat loosely confined in the within the 10nm thin ITO, which is in close proximity from the metal slot arms, only separated by the 5 nm gate oxide (i, Figure 3a). This is due to the index change in the activated ITO, which forms sufficient contrast with both the bulk ITO and metal on either sides, effectively creating a Metal-Oxide-Semiconductor (MOS) capacitor facilitating electromagnetic charge storage. This effect occurs for up to near the epsilon-near-zero (ENZ) point in carrier concentration leading to highest modal overlap with the ITO (ii, Figure 3a). Beyond the ENZ point, the mode starts to become more unbounded and confinement decreases as the effective index increases and the field tends to be squeezed into the lower index oxide again. In (ii), it is noticeable that almost all the light in the structure is compressed inside the activated ITO accumulation layer of about 10 nm thickness. A similar trend is observed for the hybrid ITO case. However the field confinement of the simple hybrid waveguide (iii, Figure 3a) offers a higher overlap for the ON-state. This however is necessarily desired since the ITO has usually a higher loss than the Silicon (typical SOI) part of the hybrid design. Hence lower overlap factor for the modulator ON state would be beneficial to reduce the insertion loss.

For the hybrid ITO case, (iii) is the starting point in the range of carrier concentration for ITO. The most confined state is shown in (iv), which has a carrier concentration level near ENZ. Most light is confined in ITO in this state thus light–matter interaction in ITO is at its maximum level here. By means of the effective optical capacitance in the gap (oxide + ITO), arising from the continuity of the displacement field at the material interfaces, the light is strongly confined in the gap region in the hybrid structure.[57] At the start of the carrier concentration range, ITO index is more than that of the oxide and light being prone to occupy spaces with lower index, the light in the mode tends to be more squeezed in the oxide (Figure 3a, iii), resulting in a lower confinement factor. With tuning, as the carrier concentration increases, the material index of ITO drops to values lower than that of the oxide. This causes a change in the illumination profile of the mode and more light now is confined in the ITO layer. (Figure 3a, iv) As a result, the confinement factor increases significantly near ENZ. This effect is notably higher near the ENZ region in carrier concentration. For carrier concentrations beyond ENZ, similar to the slot ITO case, more light, is confined into the oxide layer rather than the ITO since the oxide layer has a lower index, thus the confinement decreases. Hence, the slot ITO structure can be thought of as two hybrid modes rotated vertically and then mirrored sideways. The mode structure sideways consists of material interfaces of metal, oxide, activated ITO and bulk ITO respectively; and the same combination mirrored. The activated ITO i.e. accumulation layer of about 10 nm from the oxide interface is, in effect, similar to the ITO layer in the hybrid case. The rest of the ITO material remains in its bulk state and can be compared to the *Si* in the hybrid silicon case. In this structure, similarly the oxide and activated ITO confine more light at the starting carrier concentration. But the unactivated ITO confines considerable amount of light as well. (Figure 3a, i) This happens because of the mirror symmetry in this mode. The asymmetry in the hybrid mode causes the light to be squeezed into the gap creating an optical capacitance able to store the electromagnetic energy. But in the slot structure, the symmetry allows the light to be more spread out in the bulk (unactivated) ITO. This effect also is increased due to the fact that the bulk (unactivated) ITO index is considerably lower than that of bulk *Si*. The bulk (unactivated) ITO index being closer to the oxide index also decreases the forming of the optical capacitance effect. With increasing carrier concentration, i.e. tuning, as the activated ITO index becomes comparable to that of the oxide and lower, light is more confined into the activated ITO giving rise to the confinement factor. (Figure 3a, ii) This effect also can be observed increasingly more near ENZ in the carrier concentration. Similarly as the hybrid case, the confinement goes down for carrier concentrations beyond the ENZ region.

For bulk ITO, the effective *n* change is very low as tuning only changes the accumulation layer of only about 10 nm (Figure 4a). This is because the carrier accumulation is only about 10 nm into the bulk at the ITO–oxide interface, rest of the ITO remains in the bulk state. The minimal confinement for the bulk ITO mode also reflects this effect. The confinement increases a little towards the end of the carrier concentration range, and this accounts for the slight increase in the effective extinction coefficient in that range as the material becomes very lossy. The ITO slot provides better results in terms of both the effective index change and effective extinction coefficient change. Figure 3a shows a better confinement for this structure as well. The highest confined structure, in the ITO ones, is the hybrid one. (Figure 3a) This reflects a considerable change with tuning in both real and imaginary parts of the effective index. Both the slot and hybrid structures for ITO are quite lossy to begin with, as both of them are plasmonic modes which are more lossy than bulk mode, and with tuning they become even more and more lossy as the material becomes $\kappa$ – dominant. Traditionally with tuning, the effective index drops and the effective extinction coefficient rises due to the Kramers–Kronig relations. But with the ITO slot and hybrid cases we see the opposite towards the end of the carrier concentration range, near and beyond ENZ. A closer look into these reveal that the mode, in a way, flips about the ENZ region. Towards the left of the ENZ region, the mode is more confined in the ITO and follows the material traits, i.e. the mode becomes more lossy with the material becoming lossy. Near and beyond the ENZ region the mode is leaking toward the oxide layer as ITO becomes metallic across ENZ region, thus towards the end of the carrier concentration range the confinement factor decreases.

The graphene structures seem to be the ones with the lowest change in effective indices with tuning, both for the real and imaginary parts. But this is quite natural and expected as graphene used here is a single layered 2–D material, with a naturally low light confinement into the graphene layer, due to its miniscule 0.35 nm thickness (Figure 3b). But this essentially is what separates graphene from the rest of the bulk materials in this study; the unique properties of graphene and the drastic change in refractive index between 0.4 and 0.5 eV with an infinitesimally small extinction coefficient make it a formidable opponent as it enables operations in the nano–scale. The bulk graphene mode real and imaginary parts of the effective index remain fairly the same and follow some minor changes with tuning. The slot graphene structure effective $n$ increases a small amount and then decreases with tuning, following the material trait. The effective $\kappa$ stays almost constant and then decreases in the $n$–dominant region of the material, due to the fact that $\kappa$ of the material itself here is infinitesimally small, near zero. The confinement factor also follows the same trait, i.e. confinement is more as graphene becomes lossless. The hybrid graphene case is contrary to the slot, the effective index takes a dip towards the end of the chemical potential range. The effective extinction coefficient also takes a minor dip in that region. This can be explained by the confinement factor, which is less as the chemical potential reaches the $n$–dominant region of the material (Figure 3b). The drastic change in the index amounts to only small changes in the confinement factor because of the single layer of the material and the fact that only in–plane components of the associated electric field corresponding to it.

### 4.3. Interim Conclusions based on Material and Mode Analysis

The higher losses of deeply confined waveguide modes such as found in plasmonics waveguide design must be accompanied by strongly index-changing active materials for high modulator extinction ratios. That is, combining a low-index altering material such as Silicon with plasmon modes does not improve ER performance. This is because the parasitic loss incurred upon plasma dispersion tuning of Silicon is yet orders of magnitude lower compared to the modal loss of plasmonic waveguides. The high modal overlap for Silicon bulk waveguides results in almost similar index changes compared to plasmonic-Silicon based modes with lower overlap factors than bulk. For example, the modal overlap differ by a factor of 4 for the Silicon bulk relative to the plasmon-slot is compensated by the field enhancement of about 232 times in the plasmonic mode compared to that of the bulk mode. The strong index modulation of active TCO materials in combination with increasing the modal overlap upon plasmonic field squeezing offers high modulation index differences beneficial for electro-optic modulation. However, the electrical biasing scheme and subsequent index changing region of TCOs make them weak candidates when diffraction limited modes are used due to low overlap factors. For 2D materials like graphene, the combination of such a material with diffraction-limited modes (bulk-case) does not allow for strong modulation despite high material index potential due the miniscule field overlap. Furthermore, plasmonic modes need to be designed with care since their TM character (field normal to the metal surface) hinders in-plane field components inside the 2D material. If this is not observed, the only modulation performance boost relevant to the modulator is the plasmonic field enhancement upon mode squeezing that offers any out-of-plane component such as roughness or edges. On the other hand, slot waveguides do provide the correct field polarization for 2–D materials and allow for overlap factors with 2–D materials approaching 1%.[10] This is possible, when the plasmonic slot gap and metal height is each decreased in dimensions to ten's of nanometers.

## 4.4. Tuning properties in cavities

The effect of tuning is cavity dependent as the change in resonance can lead to corresponding losses in turn, relating back to the Kramers–Kronig relationships from the material model. These shift and loss relations are cavity dependent as we can relate the shift in resonance to the dominant longitudinal mode and the loss depends on the Q-factor of the cavity as well.

### 4.4.1. The Fabry – Pérot Cavity:

The Fabry–Pérot (FP) cavity is one of the most widely used empirical cavities with established formulae for analytical manipulations to aid our calculations.[18]

$$Q = -\frac{2\pi}{\lambda} \frac{2n_{eff}L}{\log[R_1 R_2 (1-T_{loss})^2]} \tag{14}$$

where $Q$ denotes the Q–factor of the cavity, $R_1$ and $R_2$ are the optical reflectivity of the cavity facets, $T_{loss}$ is the fractional internal loss per pass and $L$ is the length of the cavity. Silver (*Ag*) mirrors are used to form the FP cavity because the air reflection in ITO is very low and is not enough to form a strong feedback system, lacking the necessary finesse in the design overall. So the corresponding mirror reflectivity becomes

$$R_1 = R_2 = \left| \left( \frac{\tilde{n}_{Ag} - \tilde{n}_{eff}}{\tilde{n}_{Ag} + \tilde{n}_{eff}} \right)^2 \right| \qquad (15)$$

The total loss in the cavity has two major components – material loss and metallic loss. Material loss is due to the active material itself being lossy with considerable effective extinction coefficients. The use of metals for mirrors in the cavity does allow better reflectivity, but the high extinction coefficient of metals incur some inevitable loss in the system. This metallic loss is due to the fact that trapped light inside the cavity can penetrate the metal mirrors based on skin depths at respective frequencies resulting in field penetration loss in the mirrors.

$$T_{loss} = \alpha_{abs\_total} + \alpha_{pen} \qquad (16)$$

$$\alpha_{abs\_total} = 1 - e^{-\alpha_{abs}L} \qquad (17)$$

$$\alpha_{abs} = \frac{4\pi\kappa_D}{\lambda} \qquad (18)$$

$$\alpha_{pen} = \left( \frac{4\pi\kappa_m}{\lambda} \right) \delta_s \cdot \frac{2\delta_s}{2\delta_s + L} \qquad (19)$$

where $\alpha_{abs}$ is the absorption coefficient of the active material, $\kappa_D$ is the effective extinction coefficient of the active material, $\lambda$ is the operating wavelength. $\alpha_{abs\_total}$ denotes the material loss i.e. light absorption in the cavity and $\alpha_{pen}$ denotes the metallic loss due to field penetration into the mirrors. $\kappa_D$ and $\kappa_M$ are the extinction coefficients of dielectric and metal materials, respectively, $\delta_s$ is the penetration depth at which the field magnitude drops to $1/e$ of the surface value, and is given by $\delta_s = \sqrt{\lambda \varepsilon_o c / \pi \sigma}$, $\sigma$ is the electrical conductivity in $\Omega^{-1} m^{-1}$, and $\varepsilon_o$ is the vacuum permittivity.[18] $\kappa_m$ is high for metals, i.e. $Ag$ in this case; and $\delta_s$ is the skin depth or penetration depth of the metal at corresponding frequency of the trapped light. The dominant mode in the cavity is found from the FEM simulations. The corresponding longitudinal mode order in the cavity is then found by $m = 2n_{eff}L/\lambda$.[58] The effects of applied bias or tuning of this mode is found from the FEM simulations. In theory, setting the longitudinal mode order from the OFF to ON states and finding the change in the resonance, i.e., the shift in the wavelength for different states, reflects the tunability of the cavity.

$$\Delta\lambda = \frac{2L}{m} |n_{eff,ON} - n_{eff,OFF}| \qquad (20)$$

## 5. Results & Discussion

### 5.1. Electro – Optic Applications in Fabry – Pérot Cavities

As the rationale for using a cavity is to build up resonances and tune the resonances thereof, it is evident that using a feedback system such as a cavity should be intended for EO applications only. If the corresponding losses are sufficiently large that all the light injected into the cavity is absorbed along the propagation direction, there will be no formation of resonances thereof. In other words, EA operations inside a cavity system do not offer any meaningful designs, since the absorptive nature of the cavity upon biasing pulls any Q factor of the cavity to zero. However, it is perceivable that there is a cavity-material combination window over which some amount of Q could provide enhanced LMIs. Here we focus on the discussion of the EO aspects. The shift in resonance with the change in carrier concentration, i.e. tuning, $\Delta\lambda$ corresponds to the change in the effective refractive index (real part), $\Delta n_{eff}$ from the modal tuning properties section above. The loss also increases with tuning as a direct result from the Kramers–Kronig relations, and our aim is to optimize the ratio of obtainable tuning which improves ER relative to this incurred losses, i.e. $\text{FOM}_{\text{EOM-cavity}} = \Delta\lambda/\Delta\alpha$. This change in the loss, $\Delta\alpha$ is a function of the modal effective extinction coefficient change, $\Delta\kappa_{eff}$. We note that with the introduction of a cavity other losses are being added, but the fundamentally strong relationship is mainly corresponding to the material and subsequent effective modal extinction coefficients. For better EO performances, it is typically desired that the tuning aspect should outweigh the losses concerned. This essentially means that we want to maximize $\Delta\lambda$ without making $\Delta\alpha$ too large. It is clear that we have to encounter the rising losses as a result of tuning, but we want to minimize this effect as much as possible for these cases. So we devise a figure of merit for these kinds of devices with the ratio $\Delta\lambda/\Delta\alpha$. The $\Delta\lambda/\Delta\alpha$ ratio for all the cavities are calculated and plotted with respect to the key tuning parameter changes, i.e. carrier concentration for *Si* and ITO structures and chemical potential for the graphene structures on the vertical axis; and scaling of the cavities on the horizontal axis (**Figure 5**) To evaluate the results in the same scale the cavity lengths are swept from 10 nm to 50 $\mu$m. The logarithm of the $\Delta\lambda/\Delta\alpha$ ratio is plotted for a better dynamic range. Here, a negative value of the ratio signifies that $\Delta\alpha$ is greater than $\Delta\lambda$ for those values. This essentially means that the losses overturn the tuning aspects for those levels, and the loss takes over in the cavity and EO modulation is limited. The material models suggest monotonically decreasing functions for the real parts of the index, *n* for *Si* and ITO. The graphene *n* is also decreasing between 0.4 eV to 0.5 eV. This necessarily means that the tuning of resonances can only shift the resonances to the left of the resonance at the start point in tuning. In other words, tuning corresponds to blue–shifts in resonance. We investigate the scaling aspects of the cavities also as the Q–factors are cavity length dependent. Typically with low loss, higher cavity length yields higher Q–factors corresponding to better performances. This automatically puts the bulk structures ahead because of their low effective extinction coefficients. However, the loss term, $\Delta\alpha$ is scaling dependent whereas the shift in resonance $\Delta\lambda$ does not depend on the length of the cavity. The modal aspects of tuning can be observed in the vertical axes of all of the plots in figure 5, and the scaling aspects, on the other hand, can be observed in the horizontal axes. In regards to the mode structures, it is noticeable in the first look that the bulk structures are the most favorable ones for tuning. If scaling is not a factor in consideration, the bulk structures can offer good results. However, since the cavity length is proportional to the electrical capacitance and energy consumption (joules-per-bit), a shorter cavity size is desired. The bulk graphene case shows high performance, with the bulk ITO structure in a close second position. But if we upscale to longer cavity lengths, the *Si* bulk structure can outperform the ITO bulk one. This happens due to the inherent higher extinction coefficient of the ITO material; the $\Delta\lambda/\Delta\alpha$ ratio starts to decrease after a certain cavity length (near $200\mu$m) for the bulk ITO structure, whereas the ratio keeps increasing for the bulk *Si* one. Hence if footprint is not a factor, a cavity with bulk *Si* can fetch higher tuning than the bulk ITO one at higher cavity lengths. The bulk *Si* mode exhibits quite a change in the effective extinction coefficient (figure 3b), but this feature is not noticeable in figure 5, because even with a high change, the effective imaginary part is still very low. Q for bulk *Si* cavity design is sufficiently larger with tuning compared to the others. (**Figure S3**) So the higher Q–factor outweighs the imaginary part of the effective index change effects.

Scaling also plays a vital role in the tuning performances in cavities. All the structures exhibit little to no tuning for lower scaling. This is due to the metal mirror feedback systems of the cavities. For shorter cavity lengths, the Fabry–Pérot structure effectively becomes a metal-insulator-metal (MIM) structure causing the Q–factor of the cavity to plummet. Also at higher scaling, all the mode structures with sufficient modal loss to begin with, i.e. the more confined ones (except the bulk ones), exhibit a descend in the $\Delta\lambda/\Delta\alpha$ ratio. This, in turn, reflects the fact that the more confined the mode is, the shorter the device can be. For the more confined structures (slot and hybrid), around micron level scaling proves to be optimal (Figure 5). Even with optimal scaling, it is noticeable that the ITO and graphene structures outperform the *Si* ones mainly because of the material traits following steeper index change. Near the micron scale, the slot structure is better than the hybrid structure for EO operations, and suggests high-performance micrometer small mach-zender-interforemeter (MZI) based EO modulators. The *Si* slot is about 11.5 times better than its hybrid counterpart, the ITO slot is about 1.54 times better than its hybrid counterpart and the graphene slot is about 1.39 times better than its hybrid counterpart; all for the highest possible tuning that correspond to the highest values in the $\Delta\lambda/\Delta\alpha$ ratio near the micron level scaling. In the ITO plasmonic structures, i.e. the slot and hybrid, it is noticeable that similar performance devices can be obtained with a reduced footprint of about an order of magnitude by employing near ENZ modulation off state. The slot structure can fetch about 1 order of magnitude reduced scaling for similar performance in the optimal scaling region, and the HPP structure can fetch about 1.5 orders of magnitude reduced scaling similarly. If the length is fixed near the micron scale, we can compare the different material based cavity performance with the same mode structures, for example slot *Si* with slot graphene. First for the bulk structures, the graphene and ITO ones are nearly 43 and 7 times better, respectively, than the *Si* one, in terms of the ratio, for our intended purpose. Similarly for the slot structures, graphene and ITO ones are nearly 49 and 11 times better, respectively than their *Si* counterpart. And finally for the hybrid ones, the graphene and ITO ones are 410 and 84 times better than the *Si* one, respectively. So overall graphene appears to be a very formidable opponent to the bulk materials. Within the bulk materials, ITO outshines *Si* in more confined and shorter structures.

## 5.2. Electro – Absorptive Linear Performance

The cavity based electro–optic phase modulator is a crucial component in the on chip photonic network, however, the cavity size varies but is usually large compared to other on chip devices, which limits the on chip density when optimizing the network performance. Compared to phase modulation by using cavities, electro–absorptive modulators, by tuning the loss term of active materials, provide a rather compact device footprint, giving ways to further increase the on chip bit area density. In this work, we numerically analyzed the EA property given the aforementioned materials and optical modes. The performance is characterized by extinction ratio (ER) in dB form (**Table 1**). Those results are analytically calculated from the effective mode index retrieved from FEM analyses. For *Si* modes, as expected, the ER for all three modes are low due to the fact that the imaginary part of refractive index changes insignificantly during carrier injection. Similarly, although graphene has a drastic refractive index change when tuning chemical potential, the EA performance for graphene is rather poor, mainly because graphene modes suffer from the low confinement given the single atom layer thickness. ITO has a modest index change compared to *Si* and graphene, however, has the strongest EA response. For ITO bulk mode, as the confinement factor is quite low which is depicted in Figure 4, the extinction changes slightly between on and off states. With the help of plasmonic modes, most portion of light could be squeezed into the active layer of ITO (Figure 4), which leverage the ER to a much higher level, especially in hybrid mode, the ER is as high as 2.9 dB/$\mu m$.

## 6. Conclusion

Out of the three materials chosen for this study, ITO and Graphene clearly outperform conventional *Si,* in both elctro–optic and electro–absorptive cases. Also the different modes discussed have their distinguishable advantages. The more confined slot and hybrid modes lead to shorter device lengths and are beneficial for acheiving improved results for small scaling. But for overall better electro–optic results the bulk structure is the favorable option. If the scaling or footprint is not a deciding factor, one can easily employ the bulk structure to get enhanced modulation strength using cavity feedback. However, scaling and footprint reduction is of utmost importance to enabling nanoscale operation and improving the on-chip packaging density. The unique elctro-optice property of graphene with the sharp real part of the index change and the almost constant infinitesimal imaginary part of the index in that region, can make it the highest performing material in terms of modulation strength. But consideration must be given to the corresponding confinement factor and to placing the Graphene sheet in such a manner that the in-plane components of the involved electric field in the mode is maximized. In our choice of the mode structures, the graphene slot served these conditions better; and that is why it was the best performing EOM cavity in the plasmonic ones. Also employing near ENZ modulator off state in the ITO plasmonic structures can benefit cavity optimal scaling by about an order of magnitude. This essentially means similar performance cavity modulators in a reduced footprint of about an order of magnitude. For EAM operations, the more lossy plasmonic structures should be considered. ITO slot and HPP structures can offer superior performance, whereas the graphene slot is also noteworthy considering the atomic thin layer of graphene. Depending on the desired effects and scaling concerns, the proper material – modal combinations should be employed.

## Acknowledgements

This work was supported by the Army Research Office under the contract number W911NF-16-2-0194.

# Figures

**Figure 1.** Material refractive indices and extinction coefficients vs. wavelength dispersion by varying carrier concentration, $N_c$ for $Si$ and ITO, and chemical potential, $\mu_c$ for Graphene (a – f). $\delta n/\delta N_c$, $\delta \kappa/\delta N_c$ and $\delta n/\delta \kappa$ vs. carrier concentration, $N_c$ (for $Si$ & ITO) and chemical potential, $\mu_c$ (for Graphene) at $\lambda = 1550\ nm$ (g – l). The n–k vs. chemical potential, $\mu_c$ for Grphene is shown as an inset of the Graphene $\delta n/\delta \kappa$ plot (l).

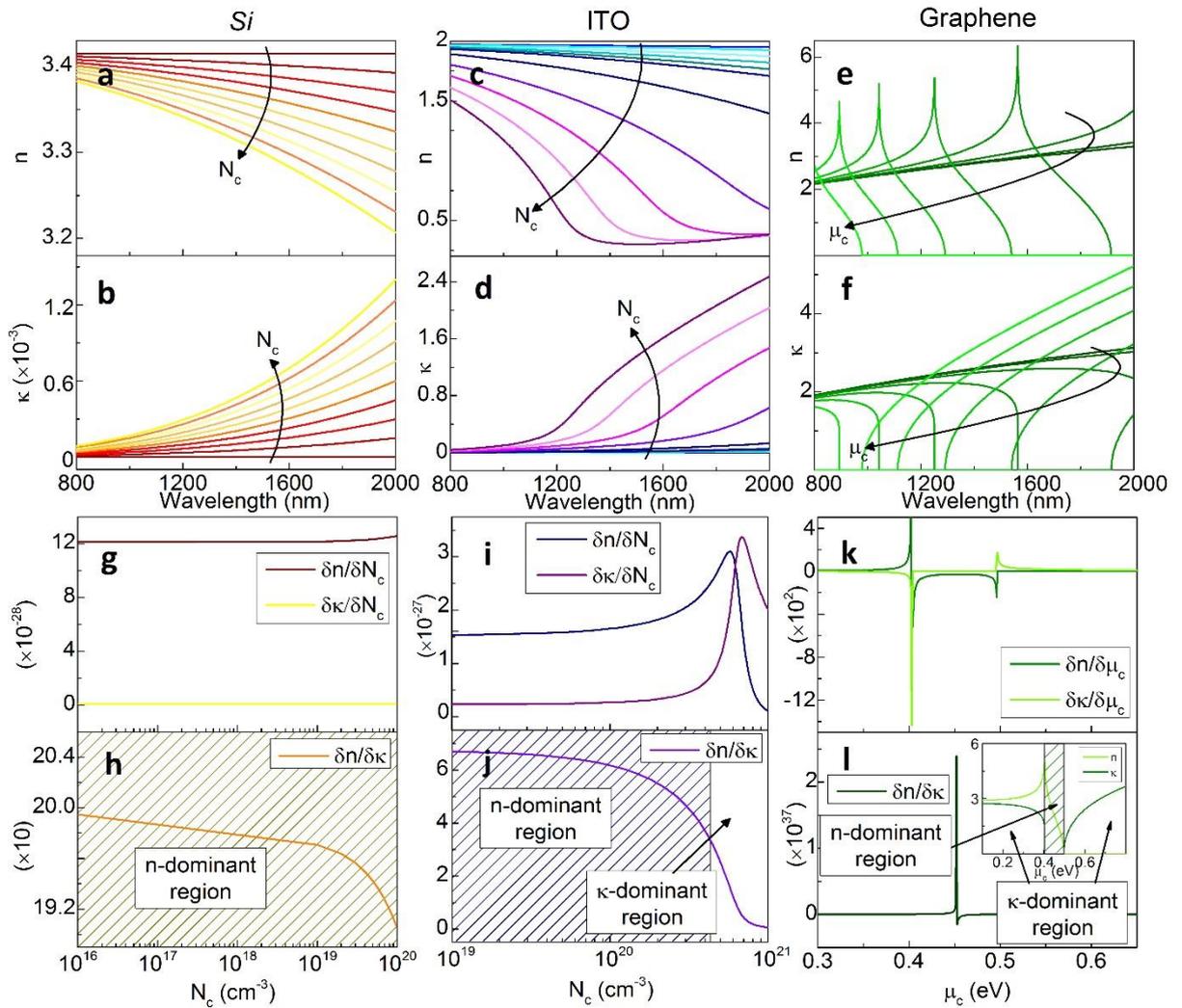

**Figure 2.** Schematic of the mode structures and FEM simulated modes for all the structures at their respective starting point from the material dipersion at $\lambda = 1550\ nm$. $\lambda/n_{\text{Si}} = 451\ nm$, $\lambda/n_{ITO} = 800\ nm$, $h_{Si} = h_{ITO} = 200\ nm$, $h_{ITO_{blk}} = 600\ nm$, $h_{slot} = 100\ nm$, $w_{slot} = 300\ nm$, $h_{Si_{slot}} = 30\ nm$, $g = 20\ nm$, $w_{metal} = 550\ nm$, $w_{ITO} = 300\ nm$, $h_{metal} = 20\ nm$, $h_{ITO_{hyb}} = 10\ nm$, $w_{ITO_{hyb}} = 250\ nm$. The simulated results are shown in log scale due to their largely varying electric field strengths. All gate oxides in this work have thickness $h_{ox} = 5\ nm$ to ensure similar electrostatics.

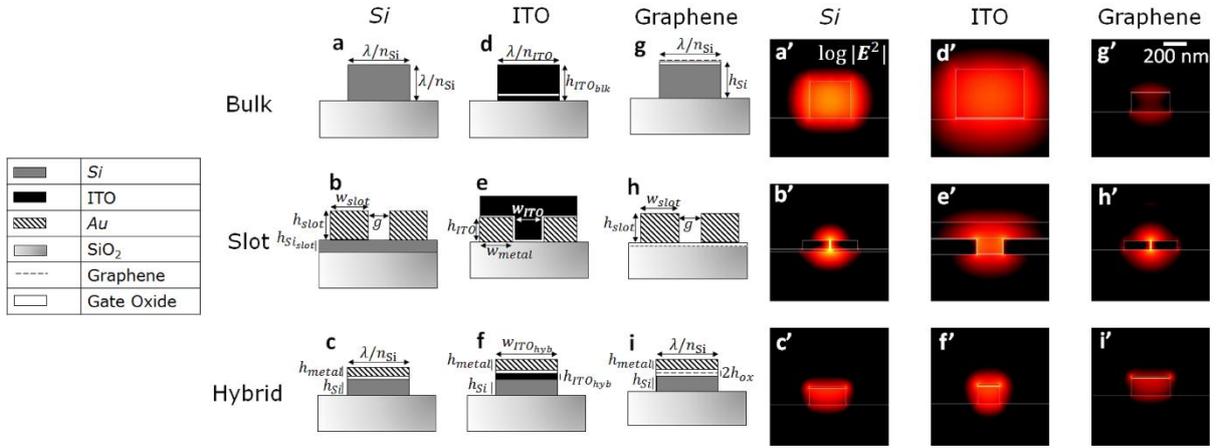

**Figure 3.** Confinement factors of the modes from figure 2 with tuning. Confinement factors corresponding to the *Si* and ITO modes vs. carrier concentration (a) and confinement factors corresponding to the Graphene modes vs. chemical potential (b). (i, ii) ITO slot at $10^{19}$cm$^{-3}$ and $6\times10^{20}$ cm$^{-3}$, (iii, iv) ITO hybrid at $10^{19}$ cm$^{-3}$ and $6\times10^{20}$ cm$^{-3}$; respectively.

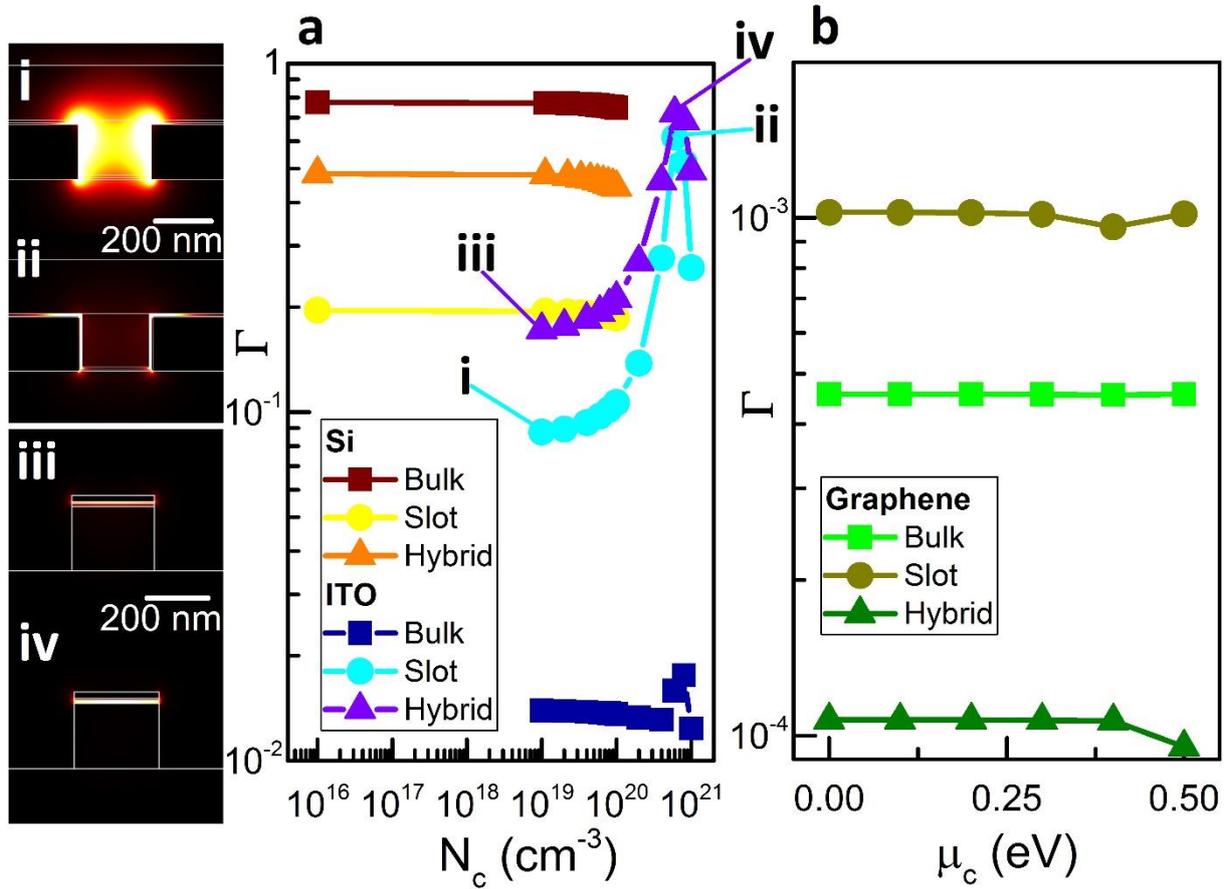

**Figure 4.** Simulated results for effective indices for all the modes in Figure 2 for different applied bias. (a) Real part of the effective refractive index for *Si* and ITO vs. carrier concentration, (b) Imaginary part of the effective index i.e. effective extinction coefficient for *Si* and ITO vs. carrier concentration, and (c) Real and imaginary parts of the effective index for graphene vs. chemical potential.

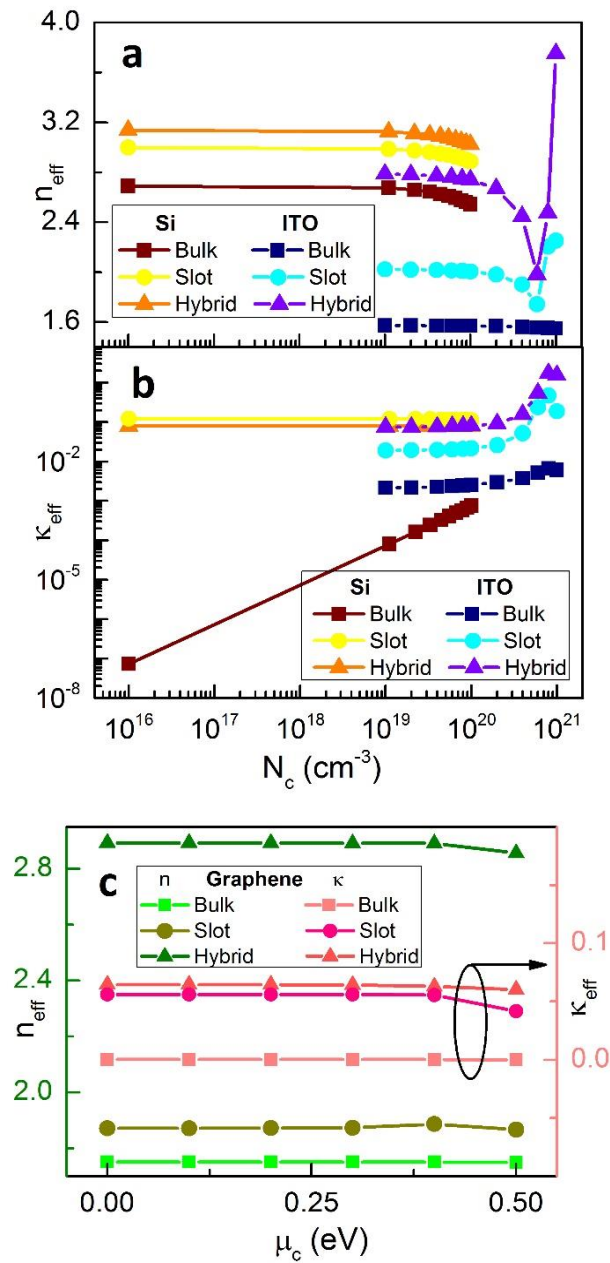

**Figure 5.** Fabry-Perot based cavity resonance shift-loss-ratio, i.e. $\log(\Delta\lambda/\Delta\alpha)$, vs. carrier concentration levels in cm$^{-3}$ (for *Si* and ITO) and chemical potential in eV (for Graphene) vs. scaling ($\mu$m) for all 9 cavities based on the modes from Figure 2. Every point on the y –axes on these graphs represent corresponding tuning from the start point in either carrier concentration (a – f) or chemical potential (g – i) to that point. Each color level in the color bar correspond to a change of approximately $2.27 \times 10^4$ cm$^2$ in the actual $\Delta\lambda/\Delta\alpha$ ratio.

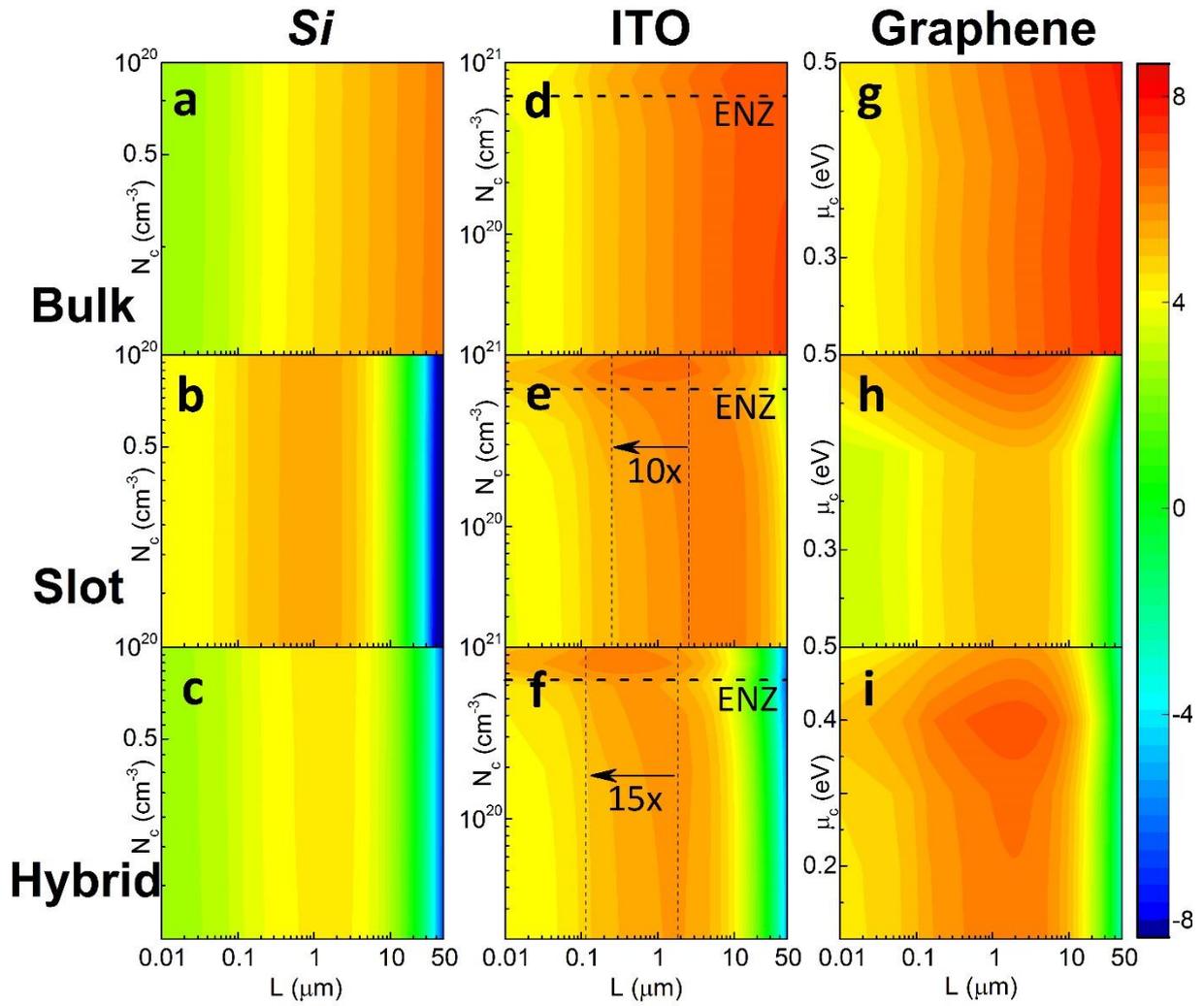

**Table 1.** Table of Electro – Absorptive Performances Characterized by Extinction Ratio (ER) for All the Structures

| Material / Mode [dB/$\mu m$] | Bulk | Slot | Hybrid |
|---|---|---|---|
| *Si* | 0.027 | 0.372 | 0.022 |
| ITO | 0.058 | 1.155 | 2.928 |
| Graphene | 0.004 | 0.505 | 0.148 |